\documentclass[11pt,english]{revtex4}
\usepackage{lmodern}

\usepackage[T1]{fontenc}
\usepackage[latin9]{inputenc}
\usepackage{color}
\usepackage{babel}

\usepackage{amsmath}
\usepackage{amssymb}
\usepackage{esint}
\usepackage[unicode=true, pdfusetitle,
 bookmarks=true,bookmarksnumbered=false,bookmarksopen=false,
 breaklinks=false,pdfborder={0 0 1},backref=false,colorlinks=false]
 {hyperref}

\makeatletter

\providecommand{\tabularnewline}{\\}

\@ifundefined{textcolor}{}
{%
 \definecolor{BLACK}{gray}{0}
 \definecolor{WHITE}{gray}{1}
 \definecolor{RED}{rgb}{1,0,0}
 \definecolor{GREEN}{rgb}{0,1,0}
 \definecolor{BLUE}{rgb}{0,0,1}
 \definecolor{CYAN}{cmyk}{1,0,0,0}
 \definecolor{MAGENTA}{cmyk}{0,1,0,0}
 \definecolor{YELLOW}{cmyk}{0,0,1,0}
 }

\usepackage{latexsym}\usepackage{bm}

\makeatother

\makeatother

\begin{document}

\title{All Bulk and Boundary Unitary Cubic Curvature Theories in Three Dimensions}

\author{\.{I}brahim Güllü }

\email{e075555@metu.edu.tr}

\affiliation{Department of Physics,\\
 Middle East Technical University, 06531, Ankara, Turkey}

\author{Tahsin Ça\u{g}r\i{} \c{S}i\c{s}man}

\email{sisman@metu.edu.tr}

\affiliation{Department of Physics,\\
 Middle East Technical University, 06531, Ankara, Turkey}

\author{Bayram Tekin}

\email{btekin@metu.edu.tr}

\affiliation{Department of Physics,\\
 Middle East Technical University, 06531, Ankara, Turkey}

\date{\today}
\begin{abstract}
We construct all the bulk and boundary unitary cubic curvature parity
invariant gravity theories in three dimensions in (anti)-de Sitter
spaces. For bulk unitarity, our construction is based on the principle
that the free theory of the cubic curvature theory reduces to one
of the three known unitary theories which are the cosmological Einstein-Hilbert
theory, the quadratic theory of the scalar curvature or the new massive
gravity (NMG). Bulk and boundary unitarity in NMG is in conflict;
therefore, cubic theories that are unitary both in the bulk and on
the boundary have free theories that reduce to the other two alternatives.
We also study the unitarity of the Born-Infeld extensions of NMG to
all orders in curvature.\tableofcontents{} 
\end{abstract}
\maketitle

\section{Introduction}

In three dimensions, there are three parity invariant pure gravity
theories that are known to be unitary in the sense of tachyon and
ghost freedom at the tree level. These are the (cosmological) Einstein-Hilbert
theory with no local degrees of freedom, the quadratic theory built
from the curvature scalar with the Lagrangian density $R-2\Lambda_{0}+aR^{2}$
which has a single massive scalar degree of freedom %
\footnote{In fact, pure $f\left(R\right)$ theory with the Lagrangian density
$R+\alpha R^{n}$ is also unitary, but for the discussion in this
paper $n=2$ theory is relevant.%
}, and the new massive gravity (NMG) defined by the action \cite{BHT1,BHT2}\begin{equation}
I=\frac{1}{\kappa^{2}}\int d^{3}x\,\sqrt{-g}\left[\sigma R-2\lambda_{0}m^{2}+\frac{1}{m^{2}}\left(R_{\mu\nu}^{2}-\frac{3}{8}R^{2}\right)\right],\label{eq:nmg_action}\end{equation}
 that provides a nonlinear extension of the Pauli-Fierz massive spin-2
theory with two degrees of freedom. Here, $\sigma=\pm1$ or it could
be set to zero to obtain a purely quadratic theory. The important
point is that, with some constraints on the parameters, these three
theories exhaust the list of unitary pure gravity theories in (anti)-de
Sitter {[}(A)dS{]} and flat spaces in three dimensions. Therefore,
if one searches for a unitary theory built from arbitrary powers of
the Ricci scalar and the tensor, then the propagator of that theory
should reduce to one of these unitary theories, with possibly redefined
parameters (such as mass, cosmological constant etc.). In flat backgrounds,
the problem is trivial: Any higher derivative (cubic and more) deformation
of the above theories is allowed since the propagators are intact
in this background. But, in constant curvature backgrounds, which
we shall deal with in this paper, generically, all the higher derivative
terms contribute to the propagators and therefore the unitarity analysis
is actually quite involved. However, as we shall show in detail, tree-level
unitary theories can be constructed systematically by studying their
propagators with the recently developed tools in \cite{Gullu_UniBI}
and with the earlier tools of \cite{Hindawi} for analyzing the unitarity
of a higher derivative theory around (A)dS backgrounds. \textcolor{black}{In
general, there are several motivations for introducing higher powers
of curvature tensors in a gravity theory. First, string theory requires
higher curvature corrections; for example, cubic curvature corrections
are given in \cite{Tseytlin}. Secondly, in four dimensions, asymptotic
safety approach to quantum gravity (see \cite{Niedermaier} for a
review) involves contributions of the every possible term constructed
by curvature tensors that is consistent with general covariance. Hence,
in the effective field theory perspective, Einstein's gravity which
is nonrenormalizable should be augmented with higher curvature terms
obeying the symmetry of the theory. An efficient way of analyzing
the effects of these higher curvature terms on the propagator structure,
and consequently on the unitarity of the theory is considered in this
paper.} In fact, as an example, we will construct all the unitary
theories in three dimensions that are built from at most the cubic
powers of the Ricci tensor.

Several extensions of NMG have already appeared recently: In \cite{Sinha},
cubic and quartic extensions of NMG was found using the requirement
that a \emph{simple} (essentially integrable) holographic $c$-function
exists. In \cite{GulluBINMG,Gullu_cfuncion}, a Born-Infeld (BI) type
action was defined which extends NMG up to any desired order in the
curvature (and in particular reproduces the same cubic and quartic
extensions of \cite{Sinha} with fixed parameters at each order of
the curvature) and which has a holographic $c$-function. In \cite{Paulos},
order by order extension of NMG was introduced again using the notion
of a holographic $c$-function. This order by order extension also
matches the curvature expansion of the Born-Infeld extended NMG \cite{Gullu_cfuncion}. 

It is worth to stress again, in constructing a generic unitary theory
at any powers of curvature, our main principle is the following: \emph{The
propagator of the theory should reduce to the propagator of the known
three unitary parity invariant theories after possible redefinitions
of the parameters.} Note that this principle is merely a restatement
of the unitary extension of a theory and does not assume any strong
conditions such as the existence of a simple holographic $c$-function
or the condition that the resulting theory can be obtained from a
BI-type action.

Up to now, we have discussed bulk unitarity only. For AdS spaces,
unitarity on the boundary is also an important issue because of the
AdS/CFT correspondence. Out of the three bulk unitary theories, NMG
always gives a nonunitary theory on the boundary \cite{BHT2}. The
other two theories have rather wide ranges of the parameters which
allow both bulk and boundary unitarity. Therefore, in AdS, if a cubic
theory is unitary in the bulk and on the boundary, then its free theory
reduces to either cosmological Einstein-Hilbert or the $R-2\Lambda_{0}+aR^{2}$
theory.

The cubic theory found before \cite{Sinha,GulluBINMG} is a single
member of the continuous family of bulk unitary theories that we shall
present. Moreover, we will more directly show the region where this
cubic theory is unitary. In principle, our analysis can be extended
to any powers of curvature tensors and to any dimensions. We will
also give two examples of arbitrary power theories: The so called
Born-Infeld extension of new massive gravity and its close cousin
\cite{GulluBINMG}, specifically we will show that their propagators
reduce to that of NMG. Namely, like the cubic theory found in \cite{Sinha},
BINMG is unitary in the bulk only. 

Since NMG (\ref{eq:nmg_action}) plays an important role in the construction
of cubic or higher order theories, let us recapitulate its properties.
For proper ranges (which we shall discuss) of the dimensionless parameters
$\sigma$, $\lambda_{0}$ and the dimensionful parameter $m^{2}$,
NMG is a tree-level (bulk) unitary theory generically describing a
massive spin-2 excitation with mass $M^{2}=\left(-\sigma+\frac{\lambda}{2}\right)m^{2}$
at the linearized level around both flat and (A)dS backgrounds \cite{BHT1,Nakasone,Deser,Gullu1,BHT2,Gullu2,Blagojevic}.
Here, the effective cosmological constant is $\Lambda=\lambda m^{2}$
with $\lambda=-2\left(\sigma\pm\sqrt{1+\lambda_{0}}\right)$. In what
follows, we will work with the mostly plus signature, assume $\kappa^{2}>0$,
and our convention for the sign of the Riemann tensor follows from
$\left[\nabla_{\mu},\nabla_{\nu}\right]V^{\sigma}\equiv R_{\mu\nu\phantom{\sigma}\rho}^{\phantom{\mu\nu}\sigma}V^{\rho}$.
In flat backgrounds, unitarity analysis of this model is quite straightforward
and has been carried out in several places, but in (A)dS backgrounds
the analysis is somewhat more complicated: In \cite{BHT2}, the theory
was shown to be formally equivalent to the Pauli-Fierz massive gravity
in (A)dS, and in \cite{Gullu2} direct gauge-invariant canonical analysis
was carried out by decomposing the spin-2 field in its irreducible
parts under the rotation group.

The layout of the paper is as follows: In Section II, we start with
the most general cubic action based on the Ricci tensor and the scalar,
and find the equivalent quadratic action which has the same $O\left(h^{2}\right)$
expansion, that is the expansion in metric perturbation, as the original
cubic action. In Section III, we discussed the unitarity of Born-Infeld
extensions of NMG. In the Appendix, we explicitly calculate the $O\left(h^{2}\right)$
expansion of BINMG.

\section{Unitary Cubic Theories}

The most general cubic curvature theory built from the Ricci tensor
and the scalar is \begin{align}
I=\frac{1}{\kappa^{2}}\int d^{3}x\,\sqrt{-g} & \Biggl[\sigma R-2\lambda_{0}m^{2}+\frac{\omega}{m^{2}}\left(R_{\mu\nu}^{2}-\frac{3}{8}R^{2}\right)+\frac{\eta}{8m^{2}}R^{2}\label{eq:R3_action}\\
 & +\frac{\alpha}{6m^{4}}\left(R^{\mu\nu}R_{\nu}^{\phantom{\nu}\alpha}R_{\alpha\mu}+\beta RR_{\mu\nu}^{2}+\gamma R^{3}\right)\Biggr],\nonumber \end{align}
 where $\sigma$, $\lambda_{0}$, $\omega$, $\eta$, $\alpha$, $\beta$
and $\gamma$ are dimensionless parameters whose signs and numerical
values are arbitrary at this stage except, we normalize $\sigma^{2}=1$,
and $\omega^{2}=1$ or $\omega=0$. On the other hand, $m^{2}$ is
of $\left[\text{Mass}\right]^{2}$ dimension and without loss of generality
we choose $m^{2}>0$ and $\kappa^{2}>0$. In flat backgrounds, which
necessarily requires $\lambda_{0}=0$, we know that for \emph{any}
$\alpha$ the theory is unitary only if $\omega\eta=0$. For $\omega=0$,
the theory should have the {}``right'' sign Einstein-Hilbert term
with $\sigma=+1$. Furthermore, if $\eta$ is also set to zero in
this case, then there is no propagating degree of freedom; while for
$\eta\ne0$ there is a spin-0 excitation with mass $m_{s}^{2}\equiv\frac{m^{2}}{\eta}>0$
in order to have a nontachyonic behavior \cite{Gullu1,Gullu2}. For
$\eta=0$ and $\omega\ne0$, NMG is recovered for $\sigma=-1$ with
two spin-2 degrees of freedom having mass $m_{g}^{2}=\frac{m^{2}}{\omega}$
with $\omega>0$ \cite{BHT1}. We will not consider the case when
$\sigma=0$. Therefore, in flat space, the already known picture at
the quadratic level does not change at the cubic or higher levels.
Thus, the main question is to find possible ranges of these parameters
for which this theory is unitary around its \emph{constant curvature}
vacua. To answer this question, one has to find the $O\left(h_{\mu\nu}^{2}\right)$
action where $h_{\mu\nu}\equiv g_{\mu\nu}-\bar{g}_{\mu\nu}$ and $\bar{g}_{\mu\nu}$
is the (A)dS vacuum (or vacua) for which $\bar{R}_{\mu\nu}=2\lambda m^{2}\bar{g}_{\mu\nu}$.
One can directly compute the $O\left(h_{\mu\nu}^{2}\right)$ action
of (\ref{eq:R3_action}), but this is highly tedious and such a direct
approach would be practically impossible for some arbitrary $R^{n}$
theories. Therefore, we will instead employ a technique developed
in \cite{Hindawi} which boils down to finding an equivalent quadratic
action which has the same propagator and the same vacua. The procedure
is quite effective and at no point one needs the complicated equations
of motion. For more details and uses of this technique see \cite{Gullu_UniBI}.
Let us now first find the maximally symmetric vacuum or vacua of (\ref{eq:R3_action}).
This can be done with the help of the \emph{equivalent quadratic}
action, as we just said, but in a simpler way the vacuum can also
be found from an \emph{equivalent linear} theory. This follows from\begin{align}
\int d^{3}x\,\mathcal{L}\left(R,R_{\mu\nu}\right)= & \int d^{3}x\,\mathcal{L}\left(\bar{R},\bar{R}_{\mu\nu}\right)+\int d^{3}x\,\left[\frac{\delta\mathcal{L}}{\delta g^{\mu\nu}}\right]_{\bar{g}_{\mu\nu}}\delta g^{\mu\nu}\label{eq:Standard_expansion}\\
 & +\frac{1}{2}\int d^{3}x\,\delta g^{\alpha\beta}\left[\frac{\delta\mathcal{L}}{\delta g^{\alpha\beta}\delta g^{\mu\nu}}\right]_{\bar{g}_{\mu\nu}}\delta g^{\mu\nu}+\dots,\nonumber \end{align}
 where $\mathcal{L}\equiv\sqrt{-g}f\left(R,R_{\mu\nu}\right)$, and
by equivalent linear action we mean an action which has the same $O\left(h^{0}\right)$
and $O\left(h\right)$ expansions as (\ref{eq:Standard_expansion}),
and equivalent quadratic action has the same $O\left(h^{0}\right)$,
$O\left(h\right)$ and $O\left(h^{2}\right)$ expansions as given
in (\ref{eq:Standard_expansion}). To find the equivalent linear or
quadratic actions, $f\left(R,R_{\mu\nu}\right)$ should be expanded
to linear (or quadratic) order in the curvature around $\left(\bar{R},\bar{R}_{\mu\nu}\right)$.
The important point is that from the linear (or quadratic) expansion
in curvature one gets all the $O\left(h\right)$ {[}or $O\left(h^{2}\right)$
{]} terms of $f\left(R,R_{\mu\nu}\right)$. Therefore, the expansion
in small curvature is not an approximation as far as the vacuum and
the propagator of the full theory is considered. {[}In these expansions
one has to keep in mind that $O\left(h^{n}\right)$ terms come from
the $\sum_{i=0}^{n}\left(R-\bar{R}\right)^{i}$ expansions.{]}

We can now start our computation and find the vacua of (\ref{eq:R3_action}).
One further simplification is to consider the Lagrangian density as
a function of $R_{\nu}^{\mu}$, in order not to introduce the metric
or its inverse during the expansion. Therefore, we have\begin{align}
f\left(R_{\nu}^{\mu}\right)\equiv & \sigma\delta_{\mu}^{\nu}R_{\nu}^{\mu}-2\lambda_{0}m^{2}+\frac{\omega}{m^{2}}\left(R_{\nu}^{\mu}R_{\mu}^{\nu}-\frac{3}{8}R^{2}\right)+\frac{\eta}{8m^{2}}\left(\delta_{\mu}^{\nu}R_{\nu}^{\mu}\right)^{2}\nonumber \\
 & +\frac{\alpha}{6m^{4}}\left[R_{\rho}^{\mu}R_{\nu}^{\rho}R_{\mu}^{\nu}+\beta\left(\delta_{\lambda}^{\gamma}R_{\gamma}^{\lambda}\right)\left(R_{\nu}^{\mu}R_{\mu}^{\nu}\right)+\gamma\left(\delta_{\mu}^{\nu}R_{\nu}^{\mu}\right)^{3}\right].\end{align}
 Then, expanding $f\left(R_{\nu}^{\mu}\right)$ to the first order
around the yet to be found background $\left(\bar{R}_{\nu}^{\mu}=2\lambda m^{2}\delta_{\nu}^{\mu}\right)$
with the assumption of small fluctuations {[}that is $\left(R_{\beta}^{\alpha}-\bar{R}_{\beta}^{\alpha}\right)$
being small{]} as \begin{equation}
f\left(R_{\nu}^{\mu}\right)=f\left(\bar{R}_{\nu}^{\mu}\right)+\left[\frac{\partial f}{\partial R_{\beta}^{\alpha}}\right]_{\left(\bar{R}_{\nu}^{\mu}\right)}\left(R_{\beta}^{\alpha}-\bar{R}_{\beta}^{\alpha}\right)+O\left[\left(R_{\beta}^{\alpha}-\bar{R}_{\beta}^{\alpha}\right)^{2}\right],\label{eq:First_order_in_R-Rbar}\end{equation}
 one obtains the equivalent linear Lagrangian density $g_{\text{lin-equal}}\left(R_{\nu}^{\mu}\right)$
after dropping the quadratic order as \begin{align}
g_{\text{lin-equal}}\left(R_{\nu}^{\mu}\right)= & \left[-2\lambda_{0}+\frac{3\lambda^{2}}{2}\left(\omega-3\eta\right)-8\alpha\lambda^{3}\left(1+3\beta+9\gamma\right)\right]m^{2}\nonumber \\
 & +\left[\sigma-\frac{\lambda}{2}\left(\omega-3\eta\right)+2\alpha\lambda^{2}\left(1+3\beta+9\gamma\right)\right]R.\end{align}
 Therefore, the equivalent linear action becomes\begin{align}
I_{\text{lin-equal}}= & \frac{1}{\kappa^{2}}\int d^{3}x\sqrt{-g}\,\left[\sigma-\frac{\lambda}{2}\left(\omega-3\eta\right)+2\alpha\lambda^{2}\left(1+3\beta+9\gamma\right)\right]\nonumber \\
 & \times\left[R-\frac{\left[4\lambda_{0}-3\left(\omega-3\eta\right)\lambda^{2}+16\alpha\lambda^{3}\left(1+3\beta+9\gamma\right)\right]}{\left[2\sigma-\lambda\left(\omega-3\eta\right)+4\alpha\lambda^{2}\left(1+3\beta+9\gamma\right)\right]}m^{2}\right].\label{eq:Cubic_lin-equal}\end{align}
 Let us stress again that (\ref{eq:Cubic_lin-equal}) and (\ref{eq:R3_action})
have the same $O\left(h^{0}\right)$ and $O\left(h\right)$ expansions.
Since $O\left(h\right)$ expansion of (\ref{eq:Cubic_lin-equal})
evaluated at $\bar{g}_{\mu\nu}$ just gives the equations of motion,
that is the Einstein tensor evaluated in the vacuum in this case,
we can easily read the vacuum, by comparing it to $\sqrt{-g}\left(R-2\lambda m^{2}\right)$
and find\begin{align}
2\lambda=\frac{4\lambda_{0}-3\left(\omega-3\eta\right)\lambda^{2}+16\alpha\lambda^{3}\left(1+3\beta+9\gamma\right)}{2\sigma-\lambda\left(\omega-3\eta\right)+4\alpha\lambda^{2}\left(1+3\beta+9\gamma\right)} & \Rightarrow4\sigma\lambda+\lambda^{2}\left(\omega-3\eta\right)-8\alpha\lambda^{3}\left(1+3\beta+9\gamma\right)=4\lambda_{0},\label{eq:First_vac_eqn_for_cubic}\end{align}
 which has always at least one real root for \emph{generic} values
of the parameters: Therefore, unlike the NMG case which requires $\lambda_{0}\ge-1$
for (A)dS to be the vacuum, for any $\lambda_{0}$ , (\ref{eq:R3_action})
has a maximally symmetric vacuum. At this stage, no restriction exists
on the ranges of the parameters, but as we will see now, unitarity
of the theory will constrain some of these parameters.

Let us now find the equivalent quadratic action by expanding $f\left(R_{\nu}^{\mu}\right)$
up to second order in the curvature:\begin{equation}
g_{\text{quad-equal}}\left(R_{\nu}^{\mu}\right)\equiv f\left(\bar{R}_{\nu}^{\mu}\right)+\left[\frac{\partial f}{\partial R_{\beta}^{\alpha}}\right]_{\bar{R}_{\nu}^{\mu}}\left(R_{\beta}^{\alpha}-\bar{R}_{\beta}^{\alpha}\right)+\frac{1}{2}\left[\frac{\partial^{2}f}{\partial R_{\sigma}^{\rho}\partial R_{\beta}^{\alpha}}\right]_{\bar{R}_{\nu}^{\mu}}\left(R_{\beta}^{\alpha}-\bar{R}_{\beta}^{\alpha}\right)\left(R_{\sigma}^{\rho}-\bar{R}_{\sigma}^{\rho}\right),\label{eq:Hindawi_exp}\end{equation}
 where\begin{align}
f\left(\bar{R}_{\nu}^{\mu}\right) & =\left[6\sigma\lambda-2\lambda_{0}-\frac{3\lambda^{2}}{2}\left(\omega-3\eta\right)+4\alpha\lambda^{3}\left(1+3\beta+9\gamma\right)\right]m^{2},\nonumber \\
\left[\frac{\partial f}{\partial R_{\beta}^{\alpha}}\right]_{\bar{R}_{\nu}^{\mu}} & =\left[\sigma-\frac{\lambda}{2}\left(\omega-3\eta\right)+2\alpha\lambda^{2}\left(1+3\beta+9\gamma\right)\right]\delta_{\alpha}^{\beta},\\
\left[\frac{\partial^{2}f}{\partial R_{\sigma}^{\rho}\partial R_{\beta}^{\alpha}}\right]_{\bar{R}_{\nu}^{\mu}} & =\frac{2}{m^{2}}\left\{ \left[\omega+\alpha\lambda\left(1+\beta\right)\right]\delta_{\rho}^{\beta}\delta_{\alpha}^{\sigma}-\frac{3}{8}\left[\omega-\frac{1}{3}\eta-\frac{8\alpha\lambda}{9}\left(2\beta+9\gamma\right)\right]\delta_{\rho}^{\sigma}\delta_{\alpha}^{\beta}\right\} .\nonumber \end{align}
 Then, collecting all these we get the equivalent quadratic Lagrangian
density \begin{align}
g_{\text{quad-equal}}\left(R_{\nu}^{\mu}\right)= & \left[-2\lambda_{0}+4\alpha\lambda^{3}\left(1+3\beta+9\gamma\right)\right]m^{2}+\left[\sigma-2\alpha\lambda^{2}\left(1+3\beta+9\gamma\right)\right]R\nonumber \\
 & +\frac{1}{m^{2}}\left[\omega+\alpha\lambda\left(1+\beta\right)\right]R_{\mu\nu}^{2}-\frac{3}{8m^{2}}\left[\omega-\frac{1}{3}\eta-\frac{8\alpha\lambda}{9}\left(2\beta+9\gamma\right)\right]R^{2},\label{eq:Equivalent_quad_Lag_of_cubic}\end{align}
 whose $O\left(h^{2}\right)$, $O\left(h\right)$ and $O\left(h^{0}\right)$
expansions match the same expansions of (\ref{eq:R3_action}). At
this stage, it is clear that there are three different ways for the
general cubic theory (\ref{eq:R3_action}) to be unitary: Its equivalent
quadratic action (\ref{eq:Equivalent_quad_Lag_of_cubic}) can be,
with redefined parameters, equal to the cosmological Einstein-Hilbert
theory or $R+aR^{2}$ theory or NMG. {[}Again, we exclude the case
for which Einstein-Hilbert term drops out.{]} First, it pays to rewrite
the equivalent quadratic action as\begin{equation}
I_{\text{quad-equal}}=\frac{1}{\kappa^{2}}\int d^{3}x\,\sqrt{-g}\left[\tilde{\sigma}R-2\tilde{\lambda}_{0}m^{2}+\frac{\tilde{\omega}}{m^{2}}\left(R_{\mu\nu}^{2}-\frac{3}{8}R^{2}\right)+\frac{\tilde{\eta}}{8m^{2}}R^{2}\right],\label{eq:Equiv_quad_act_of_cubic}\end{equation}
 where\[
\tilde{\sigma}\equiv\sigma-2\alpha\lambda^{2}\left(1+3\beta+9\gamma\right),\qquad\tilde{\lambda}_{0}\equiv\lambda_{0}-2\alpha\lambda^{3}\left(1+3\beta+9\gamma\right),\]
 \begin{equation}
\tilde{\omega}\equiv\omega+\alpha\lambda\left(1+\beta\right),\qquad\tilde{\eta}\equiv\eta+\frac{\alpha\lambda}{3}\left(9+25\beta+72\gamma\right).\label{eq:Effective_parameters}\end{equation}
 Here, it is worth restating that $\lambda$ appearing in the redefined
parameters is the vacuum of (\ref{eq:R3_action}) satisfying (\ref{eq:First_vac_eqn_for_cubic})
which can also be directly obtained by computing the vacuum of (\ref{eq:Equiv_quad_act_of_cubic})
which reads from the somewhat simpler looking expression\begin{equation}
\tilde{\sigma}\lambda+\frac{1}{4}\left(\tilde{\omega}-3\tilde{\eta}\right)\lambda^{2}=\tilde{\lambda}_{0}.\label{eq:Vacuum_of_equiv_quad}\end{equation}
 Canonical analysis of (\ref{eq:Equiv_quad_act_of_cubic}) have shown
that there are generically three, \emph{not necessarily unitary},
degrees of freedom with the masses \cite{Gullu2}:\begin{align}
m_{s}^{2} & =\left[\frac{\tilde{\sigma}}{\tilde{\eta}}-\frac{3}{2}\lambda\left(1-\frac{\tilde{\omega}}{3\tilde{\eta}}\right)\right]m^{2}\qquad\text{helicity-0 mode},\\
m_{g}^{2} & =\left[-\frac{\tilde{\sigma}}{\tilde{\omega}}+\frac{1}{2}\lambda-\frac{3}{2}\lambda\frac{\tilde{\eta}}{\tilde{\omega}}\right]m^{2}\qquad\text{helicity-}\pm\text{2 modes}.\end{align}
 For (\ref{eq:Equiv_quad_act_of_cubic}) to be unitary, the necessary
but \emph{not} \emph{sufficient} condition is $\tilde{\omega}\tilde{\eta}=0$
which again exhausts all three unitary theories. Among these theories,
NMG, for which $\tilde{\eta}=0$, seems to be the most interesting
one with spin-2 excitations (scalar mode decouples), therefore we
start with it. But, NMG in (A)dS is not unitary by default: There
are constraints on the parameters which we discuss below. Since the
parameters appear in certain combinations let us define $\xi\equiv2\alpha\left(1+3\beta+9\gamma\right)$
and $\chi\equiv\alpha\left(1+\beta\right)$, then the effective parameters
(\ref{eq:Effective_parameters}) become\[
\tilde{\sigma}\equiv\sigma-\lambda^{2}\xi,\qquad\tilde{\lambda}_{0}\equiv\lambda_{0}-\lambda^{3}\xi,\]
 \begin{equation}
\tilde{\omega}\equiv\omega+\lambda\chi,\qquad\tilde{\eta}\equiv\eta+\frac{\lambda}{3}\left(\chi+4\xi\right).\label{eq:Redefined_parameters}\end{equation}

\subsection{Reducing the cubic theory to NMG in (A)dS\label{sub:Reducing-NMG}}

Setting $\tilde{\eta}=0$, the equivalent quadratic action (\ref{eq:Equiv_quad_act_of_cubic})
reduces to NMG with $m_{g}^{2}=\left(-\frac{\tilde{\sigma}}{\tilde{\omega}}+\frac{1}{2}\lambda\right)m^{2}$
where $\lambda=-\frac{2}{\tilde{\omega}}\left(\tilde{\sigma}\pm\sqrt{\tilde{\sigma}^{2}+\tilde{\omega}\tilde{\lambda}_{0}}\right)$
which requires $\tilde{\sigma}^{2}+\tilde{\omega}\tilde{\lambda}_{0}\ge0$.
The theory is unitary if $\frac{m^{2}}{\tilde{\omega}}\left(\lambda m^{2}-2\tilde{\sigma}\frac{m^{2}}{\tilde{\omega}}\right)>0$
which comes from the ghost freedom requirement of \cite{BHT2} and
reduces to $\tilde{\omega}\lambda-2\tilde{\sigma}>0$ in our notation.
This requirement can be seen by rewriting NMG in the form of a massive
Pauli-Fierz theory at the linearized level. In de Sitter case ($\lambda>0$),
there is also the Higuchi bound \cite{Higuchi} $m_{g}^{2}\ge\lambda m^{2}$
which becomes $\frac{2\tilde{\sigma}}{\tilde{\omega}}+\lambda\le0$,
and in anti-de Sitter case ($\lambda<0$), there is the Breitenlohner-Freedman
(BF) bound \cite{BF} $m_{g}^{2}\ge\lambda m^{2}$ which is exactly
like the Higuchi bound for this three-dimensional case. {[}Strictly
speaking BF bound was derived for massive scalar field in AdS, but
it works for massive spin-2 field as well \cite{Waldron}{]} In this
setting, unitarity analysis of (\ref{eq:Equiv_quad_act_of_cubic})
for $\tilde{\eta}=0$ is the same as NMG with an essential difference:
$\tilde{\sigma}$ and $\tilde{\omega}$ are not in general $\pm1$.
However, as implied by the unitarity constraints, unitary regions
can be classified according to the signs of $\tilde{\sigma}$ and
$\tilde{\omega}$ just like in the case of NMG. Since the unitarity
regions of NMG in (A)dS were studied in detail in \cite{BHT2}, we
will not repeat the analysis here, but simply give an example in AdS
($\lambda<0$). Choose $\tilde{\sigma}<0$ and $\tilde{\omega}>0$:
BF bound is automatically satisfied, so the unique constraint on the
vacuum of the theory is $\lambda>\frac{2\tilde{\sigma}}{\tilde{\omega}}$
with $\lambda=-\frac{2}{\tilde{\omega}}\left(\tilde{\sigma}+\sqrt{\tilde{\sigma}^{2}+\tilde{\omega}\tilde{\lambda}_{0}}\right)$
which can be achieved if the parameters of the theory satisfy the
inequality \begin{equation}
0<\tilde{\lambda}_{0}<\frac{3\tilde{\sigma}^{2}}{\tilde{\omega}}.\end{equation}
 This is a rather weak condition on the parameters, therefore there
is a continuum of unitary theories. 
\begin{enumerate}
\item Choose $\sigma=-1$ and $\omega=1$: For the sake of simplicity, let
us further assume $\eta=0$ which fixes $\xi=-\frac{\chi}{4}$ that
yields $\gamma=-\frac{25\beta+9}{72}$ in terms of the original parameters
of the theory (we discuss $\eta\ne0$ cases below). Then, for $\lambda_{0}<0$
there is no unitary theory, but for $\lambda_{0}>0$ the theory is
unitary if the following conditions are met:\begin{equation}
\left(\chi\le\frac{1}{4},\quad\text{and}\quad0<\lambda_{0}<\frac{-1+\left(1-4\chi\right)^{3/2}+6\chi}{2\chi^{2}}\right)\quad\text{or}\quad\left(\chi>\frac{1}{4},\quad\text{and}\quad0<\lambda_{0}<\frac{1}{\chi}\right).\label{eq:sl0_og0_cond}\end{equation}
 For example, consider the $\chi=0$ case, it is unitary for $0<\lambda_{0}<3$
with the same vacuum as NMG, $\lambda=2\left(1-\sqrt{1+\lambda_{0}}\right)$.
In fact, NMG with $\alpha=0$ is a member of this family, since $\chi=\alpha\left(1+\beta\right)$.
But, $\beta=-1$ gives a cubic order extension which is probably the
simplest unitary one parameter extension of NMG with the action \begin{align}
I=\frac{1}{\kappa^{2}}\int d^{3}x\,\sqrt{-g} & \Biggl[-R-2\lambda_{0}m^{2}+\frac{1}{m^{2}}\left(R_{\mu\nu}^{2}-\frac{3}{8}R^{2}\right)\label{eq:One_para_ext}\\
 & +\frac{\alpha}{6m^{4}}\left(R^{\mu\nu}R_{\nu}^{\phantom{\nu}\alpha}R_{\alpha\mu}-RR_{\mu\nu}^{2}+\frac{2}{9}R^{3}\right)\Biggr],\nonumber \end{align}
 with an \emph{arbitrary} $\alpha$. The other one parameter extension
of NMG introduced in \cite{Sinha} is also a member of $\tilde{\eta}=0$
and $\eta=0$ family of unitary theories, for this case one chooses
$\beta=-9/8$ which then fixes $\gamma=17/64$ yielding an action
\begin{align}
I=\frac{1}{\kappa^{2}}\int d^{3}x\,\sqrt{-g} & \Biggl[-R-2\lambda_{0}m^{2}+\frac{1}{m^{2}}\left(R_{\mu\nu}^{2}-\frac{3}{8}R^{2}\right)\label{eq:One_para_ext_of_SG}\\
 & -\frac{4\chi}{3m^{4}}\left(R^{\mu\nu}R_{\nu}^{\phantom{\nu}\alpha}R_{\alpha\mu}-\frac{9}{8}RR_{\mu\nu}^{2}+\frac{17}{64}R^{3}\right)\Biggr],\nonumber \end{align}
 whose unitarity region is given in (\ref{eq:sl0_og0_cond}). {[}In
fact, original sign choice for $\sigma$ is $+1$ in \cite{Sinha}.{]}
Note that for $\chi=-1/2$, (\ref{eq:One_para_ext_of_SG}) reduces
to the cubic order expansion of BINMG which is unitary for $0<\lambda_{0}<-8+6\sqrt{3}$.\\
Let us also give an example for $\eta\ne0$. For simplicity choose
$\xi=0$ which yields $\eta=-\frac{\lambda\chi}{3}$, then choosing
$\lambda_{0}=1$ yields the unitarity region $-3<\chi<1$ for the
theory\begin{align}
I=\frac{1}{\kappa^{2}}\int d^{3}x\,\sqrt{-g} & \Biggl[-R-2m^{2}+\frac{1}{m^{2}}\left(R_{\mu\nu}^{2}-\frac{3}{8}R^{2}\right)-\frac{\lambda\chi}{24m^{2}}R^{2},\\
 & +\frac{\chi}{6\left(1+\beta\right)m^{4}}\left(R^{\mu\nu}R_{\nu}^{\phantom{\nu}\alpha}R_{\alpha\mu}+\beta RR_{\mu\nu}^{2}-\frac{1+3\beta}{9}R^{3}\right)\Biggr],\nonumber \end{align}
 where $\beta$ is arbitrary, and $\lambda$ is the vacuum of the
theory. Let us stress that the propagator of this theory is exactly
like NMG with redefined parameters.
\item Choose $\sigma=-1$ and $\omega=-1$: Then, $\eta=0$ theory is unitary
if $\lambda_{0}>0$ ($\lambda_{0}<0$ is ruled out) and \begin{equation}
\chi<0\quad\text{and}\quad-\frac{1}{\chi}<\lambda_{0}<\frac{1+\left(1-4\chi\right)^{3/2}-6\chi}{2\chi^{2}}.\end{equation}
 For $\eta\ne0$ and with the choice $\xi=0$, the unitary region
is $\lambda_{0}>0$ and $-\frac{3\left(\lambda_{0}+3\right)}{2\lambda_{0}^{2}}<\chi<-\frac{1}{\lambda_{0}}$.
\item Choose $\sigma=1$ and $\omega=1$: Then, $\eta=0$ theory has no
unitary region. For $\eta\ne0$, certain $\xi$ theories such as $\xi=1$
have unitary regions.
\item Choose $\sigma=1$ and $\omega=-1$: Then, $\eta=0$ theory is unitary
if \begin{equation}
-\frac{1}{4}<\chi<0\quad\text{and}\quad\frac{1}{\chi}<\lambda_{0}<\frac{1+6\chi+\left(1+4\chi\right)^{3/2}}{2\chi^{2}}.\end{equation}
 For $\eta\ne0$ and with choice $\xi=1$, the unitary region is $-2<\chi<0$
for $\lambda_{0}=1$.
\end{enumerate}
The above discussion reveals just a sample unitary cubic theories.
The other branches for various sign choices of $\tilde{\sigma}$,
$\tilde{\omega}$, $\sigma$, $\omega$ and existence or non-existence
of $\eta$ can be studied both in AdS and dS.

Although classifying all the unitary theories of the form of (\ref{eq:R3_action})
for all parameter choices is a tedious job, it is relatively easy
to find the unitary regions if some parameters are fixed as in the
cubic extension of NMG given in \cite{Sinha} and as in the case of
BINMG \cite{GulluBINMG,Gullu_cfuncion}. In \cite{Sinha}, existence
of a holographic $c$-function in a specific form is the main theme,
so in this AdS/CFT based context $\lambda_{0}$ is set to be negative
$\lambda_{0}\equiv-\frac{1}{\ell^{2}}$ and $c$-function in the considered
form can only exist, if $\beta=-9/8$ and $\gamma=17/64$ with an
arbitrary $\alpha$. Also, $\sigma=+1$ is preferred, while $\omega$
is allowed to be both $\pm1$. Then, the equivalent quadratic action
becomes\begin{equation}
I_{\text{quad-equal}}=\frac{1}{\kappa^{2}}\int d^{3}x\,\sqrt{-g}\left[\left(1-\frac{\alpha\lambda^{2}}{32}\right)R-\left(2\lambda_{0}-\frac{\alpha\lambda^{3}}{16}\right)m^{2}+\frac{1}{m^{2}}\left(\omega-\frac{\alpha\lambda}{8}\right)\left(R_{\mu\nu}^{2}-\frac{3}{8}R^{2}\right)\right],\end{equation}
 where the vacua of the theory satisfies $4\lambda+\lambda^{2}\omega-\frac{\alpha}{8}\lambda^{3}=4\lambda_{0}$.
The unitarity condition and the Higuchi/BF bounds in terms of the
original parameters of the theory become $\lambda\omega-2-\frac{\alpha\lambda^{2}}{16}>0$
and $\lambda+\frac{32-\alpha\lambda^{2}}{2\left(8\omega-\alpha\lambda\right)}\le0$,
respectively. With this setting, the theory is unitary in AdS if $\omega=+1$
and $\alpha<\frac{8}{\lambda_{0}^{2}}\left(3\lambda_{0}-8-\left(4-\lambda_{0}\right)^{3/2}\right)$;
or if $\omega=-1$, there are some constraints on $\alpha$ which
are not particularly illuminating to write. For cubic order of BINMG,
$\alpha$ is further set to be 4, but there is no unitary region for
$\sigma=+1$. On the other hand, for $\sigma=-1$ and $\omega=1$,
cubic order of BINMG is unitary in dS if $-2<\lambda_{0}<0$ and unitary
in AdS if $0<\lambda_{0}<\left(-8+6\sqrt{3}\right)$.

The above analysis shows that for nontrivial $\chi$ (or $\alpha$,
$\beta$ in terms of original parameters), there is generically a
continuous family of unitary theories, and the cubic theory of \cite{Sinha,GulluBINMG,Gullu_cfuncion}
is just an example of this family. Just like in the NMG case, there
are some special points which need further attention. For example,
at $m_{g}^{2}=\lambda m^{2}$ a new \emph{scalar }gauge invariance
of the form $\delta_{\zeta}h_{\mu\nu}=\lambda m^{2}\bar{g}_{\mu\nu}\zeta$
arises, and one has a \emph{partially massless }theory with a single
degree of freedom \cite{Nepomechie,DeserWaldron,Tekin}. {[}Note that
for the Pauli-Fierz spin-2 theory in (A)dS which is not a diffeomorphism
invariant theory, at the partially massless point the new gauge invariance
is of the form $\delta_{\zeta}h_{\mu\nu}=\nabla_{\mu}\nabla_{\nu}\zeta+\lambda m^{2}\bar{g}_{\mu\nu}\zeta$,
but the higher derivative theories that we are dealing here are diffeomorphism
invariant, and therefore, $\nabla_{\mu}\nabla_{\nu}\zeta$ part is
simply part of the diffeomorphism invariance, and should not be counted
as a new gauge symmetry.{]} The theory defined by (\ref{eq:R3_action})
has unitary partially massless \emph{regions} (in contrast to a point
in NMG) for $\chi\sigma<\frac{1}{12}$ and $\omega\lambda_{0}>-\frac{4}{3}$
(we have assumed $\eta=0$) with \begin{equation}
\lambda_{\pm}=\frac{2}{\omega}\left(-2\sigma\pm\sqrt{4+3\omega\lambda_{0}}\right).\end{equation}

Another special point is $\lambda-2\frac{\tilde{\sigma}}{\tilde{\omega}}=0$
where $m_{g}^{2}=0$ for which the linearized theory reduces to the
Proca theory for massive spin-1 field which can be seen by first writing
the equivalent quadratic action in the form of Pauli-Fierz action
by use of an auxiliary field say $f_{\mu\nu}$, and then by integrating
out the metric perturbation $h_{\mu\nu}$ which then yields a massive
spin-1 field with mass $\left(-8\frac{\tilde{\sigma}}{\tilde{\omega}}m^{2}\right)$.
The details of this procedure has been given in \cite{BHT2}. An overall
$\frac{\tilde{\omega}}{m^{2}}$ appears in the Lagrangian; therefore,
for ghost freedom $\tilde{\omega}>0$, and hence $\tilde{\sigma}<0$
is required for nontachyonic mass in the region $\sigma\chi\ge-\frac{1}{4}$
and $\omega\lambda_{0}\le4$ (we have assumed $\eta=0$) with $\lambda_{+}=4\sigma+2\sqrt{4-\lambda_{0}}$
for $\omega=1$ and $\lambda_{-}=-4\sigma+2\sqrt{4+\lambda_{0}}$
for $\omega=-1$ in both dS and AdS. {[}NMG is unitary only in AdS
for this spin-1 limit.{]}

In the above analysis, we required that the $O\left(h^{2}\right)$
theory of (\ref{eq:R3_action}) reduce to $O\left(h^{2}\right)$ of
NMG with redefined parameters. Next, we discuss the remaining two
possibilities.

\subsection{Reducing the cubic theory to Einstein's theory in (A)dS \label{sub:Reducing-EH}}

Pure Einstein's theory in three dimensions is locally trivial. Namely,
there is no propagating degree of freedom; but in any case it is a
unitary theory, and therefore the cubic theory should be allowed to
have the same $O\left(h^{2}\right)$ form as Einstein's theory around
(A)dS. This follows from (\ref{eq:Equiv_quad_act_of_cubic}) by setting
the coefficients of $R^{2}$ and $R_{\mu\nu}^{2}$ to zero. One then
obtains \begin{equation}
I=\frac{1}{\kappa^{2}}\int d^{3}x\,\sqrt{-g}\left(\tilde{\sigma}R-2\tilde{\lambda}_{0}m^{2}\right),\label{eq:Equiv_EH_act}\end{equation}
 where

\begin{equation}
\tilde{\sigma}\equiv\sigma+\frac{\lambda}{4}\left(3\eta-\omega\right),\qquad\tilde{\lambda}_{0}\equiv\lambda_{0}+\frac{\lambda^{2}}{4}\left(3\eta-\omega\right),\end{equation}
 with the vacuum $\lambda=\frac{\tilde{\lambda}_{0}}{\tilde{\sigma}}$
which reduces to $\lambda=\sigma\lambda_{0}$ (assume $\lambda_{0}\ne0$).
Then, $\beta$ and $\gamma$ can be determined in terms of other parameters
in (\ref{eq:R3_action}) as\begin{equation}
\beta=-\left(1+\frac{\omega}{\sigma\alpha\lambda_{0}}\right),\qquad\gamma=\frac{2}{9}-\frac{3\eta-25\omega}{72\sigma\alpha\lambda_{0}}.\end{equation}
 For unitarity, we should impose the right sign Einstein-Hilbert theory
that is $\sigma\left[1+\frac{\lambda_{0}}{4}\left(3\eta-\omega\right)\right]>0$.
Therefore, any cubic theory satisfying this constraint will be unitary,
yet with no local degrees of freedom at the linearized level. As a
simple example, consider $\omega=0$, $\eta=0$, then one should have
$\beta=-1$ and $\gamma=2/9$, and $\sigma=+1$ is required to have
a unitary theory with the action\begin{equation}
I=\frac{1}{\kappa^{2}}\int d^{3}x\,\sqrt{-g}\left[R-2\lambda_{0}m^{2}+\frac{\alpha}{6m^{4}}\left(R^{\mu\nu}R_{\nu}^{\phantom{\nu}\alpha}R_{\alpha\mu}-RR_{\mu\nu}^{2}+\frac{2}{9}R^{3}\right)\right].\end{equation}
 As in Sec.\ref{sub:Reducing-NMG}, the cubic theory with arbitrary
$\alpha$ and with choices $\beta=-1$ and $\gamma=2/9$ turned out
to be special. Actually, $\left(R^{\mu\nu}R_{\nu}^{\phantom{\nu}\alpha}R_{\alpha\mu}-RR_{\mu\nu}^{2}+\frac{2}{9}R^{3}\right)$
is the unique cubic curvature combination that does not effect the
free theory in both flat and (A)dS backgrounds. Let us give another
interesting example in the case for $\omega\ne0$ for which the cubic
theory \begin{align}
I=\frac{1}{\kappa^{2}}\int d^{3}x\,\sqrt{-g} & \Biggl\{\sigma R-2\lambda_{0}m^{2}+\frac{\omega}{m^{2}}\left(R_{\mu\nu}^{2}-\frac{3}{8}R^{2}\right)\label{eq:e0_o_ne_0_act}\\
 & +\frac{\alpha}{6m^{4}}\left[R^{\mu\nu}R_{\nu}^{\phantom{\nu}\alpha}R_{\alpha\mu}-\left(1+\frac{\omega}{\sigma\alpha\lambda_{0}}\right)RR_{\mu\nu}^{2}+\left(\frac{2}{9}+\frac{25\omega}{72\alpha\sigma\lambda_{0}}\right)R^{3}\right]\Biggr\},\nonumber \end{align}
 has the same $O\left(h^{0}\right)$, $O\left(h\right)$ and $O\left(h^{2}\right)$
expansions as (\ref{eq:Equiv_EH_act}). Although this theory involves
two massive excitations in flat space; in (A)dS, there is \emph{no}
propagating degree of freedom. Unitary regions of (\ref{eq:e0_o_ne_0_act})
is given in Table I.%
\begin{table}
\begin{tabular}{|c|c|c|c|c|c|}
\cline{3-6} 
\multicolumn{1}{c}{} &  & $\boldsymbol{\lambda_{0}}$ & $\boldsymbol{\sigma\left(1-\frac{\omega\lambda_{0}}{4}\right)>0}$ & $\boldsymbol{\omega}$ & \textbf{Unitary Region}\tabularnewline
\hline 
\textbf{AdS} & $\boldsymbol{\sigma=-1}$ & $\lambda_{0}>0$ & $\omega\lambda_{0}>4$ & $+1$ & $\lambda_{0}>4$\tabularnewline
\cline{2-6} 
$\boldsymbol{\sigma\lambda_{0}<0}$ & $\boldsymbol{\sigma=+1}$ & $\lambda_{0}<0$ & $\omega\lambda_{0}<4$ & \multicolumn{1}{c|}{\begin{tabular}{c}
$-1$\tabularnewline
$+1$\tabularnewline
\end{tabular}} & \begin{tabular}{c}
$-4<\lambda_{0}<0$\tabularnewline
$\lambda_{0}<0$\tabularnewline
\end{tabular}\tabularnewline
\hline 
\textbf{dS} & $\boldsymbol{\sigma=-1}$ & $\lambda_{0}<0$ & $\omega\lambda_{0}>4$ & $-1$ & $\lambda_{0}<-4$\tabularnewline
\cline{2-6} 
$\boldsymbol{\sigma\lambda_{0}>0}$ & $\boldsymbol{\sigma=+1}$ & $\lambda_{0}>0$ & $\omega\lambda_{0}<4$ & \begin{tabular}{c}
$-1$\tabularnewline
$+1$\tabularnewline
\end{tabular} & \begin{tabular}{c}
$\lambda_{0}>0$\tabularnewline
$0<\lambda_{0}<4$\tabularnewline
\end{tabular}\tabularnewline
\hline
\end{tabular}

\caption{Unitary regions for $\omega\ne0$ and $\eta=0$. }

\end{table}
 In the $\eta\ne0$ case, $\beta$ and $\gamma$ are determined as
$\beta=-1$, $\gamma=\frac{2}{9}-\frac{\eta}{24\alpha\lambda}$. To
have a unitary theory in AdS, $\eta<-\frac{4}{3\lambda_{0}}$ constraint
should be satisfied; while in dS one has $\eta>-\frac{4}{3\lambda_{0}}$.

\subsection{Reducing the cubic theory to $R-2\Lambda_{0}+aR^{2}$ theory in (A)dS\label{sub:Reducing-RaR2}}

The third and the final option of how (\ref{eq:R3_action}) can be
unitary is that it has the same propagator as the $R-2\Lambda_{0}+aR^{2}$
theory. For this to happen, the coefficient of $R_{\mu\nu}^{2}$ in
the equivalent quadratic Lagrangian density (\ref{eq:Equiv_quad_act_of_cubic})
should be set to zero. Therefore, this determines $\beta$ to be $\beta=-1-\frac{\omega}{\alpha\lambda}$.
Then, after using the vacuum equation $4\sigma\lambda+\lambda^{2}\left(25\omega-3\eta\right)+8\alpha\lambda^{3}\left(2-9\gamma\right)=4\lambda_{0}$,
or in a slightly more efficient form $4\sigma\lambda+\lambda^{2}\left(\omega-3\eta\right)-4\xi\lambda^{3}=4\lambda_{0}$,
the equivalent quadratic action can be reduced to 

\begin{equation}
I=\frac{1}{\kappa^{2}}\int d^{3}x\sqrt{-g}\,\left\{ \frac{1}{2}\left[4\sigma\lambda+\lambda^{2}\left(\omega-3\eta\right)-8\lambda_{0}\right]m^{2}+\frac{4\lambda_{0}-\lambda^{2}\left(\omega-3\eta\right)}{4\lambda}R+\frac{\sigma\lambda-\lambda_{0}}{6\lambda^{2}m^{2}}R^{2}\right\} .\label{eq:Equiv_RaR2_act}\end{equation}
 This theory is not unitary for generic values of the parameters.
One-particle amplitude \cite{Gullu1} and the canonical analyses \cite{Gullu2}
of the action \begin{equation}
I=\int d^{3}x\sqrt{-g}\,\left[\frac{1}{\kappa}\left(R-2\Lambda_{0}\right)+aR^{2}\right],\end{equation}
 show that it describes a single massive excitation with mass $m_{s}^{2}=\frac{1}{8\kappa a}-\frac{3\Lambda}{2}$
where $\Lambda$ is determined by $\Lambda-\Lambda_{0}-6a\kappa\Lambda^{2}=0$.
For unitarity $a>0$ is required for both AdS and dS, and for dS $m_{s}^{2}>0$
and for AdS we have the BF bound $m_{s}^{2}\ge\Lambda$. Therefore,
the mass of the scalar excitation described by (\ref{eq:Equiv_RaR2_act})
is\begin{equation}
m_{s}^{2}=\frac{3\lambda\left[12\lambda_{0}-8\sigma\lambda-\lambda^{2}\left(\omega-3\eta\right)\right]}{16\left(\sigma\lambda-\lambda_{0}\right)}m^{2}.\end{equation}
 The analysis of the unitary regions follows similar to Sec.\ref{sub:Reducing-NMG}
above. We will not repeat the analysis in its full detail, but just
give some examples of the regions where the cubic theory \begin{align}
I=\frac{1}{\kappa^{2}}\int d^{3}x\,\sqrt{-g} & \Biggl\{\sigma R-2\lambda_{0}m^{2}+\frac{\omega}{m^{2}}\left(R_{\mu\nu}^{2}-\frac{3}{8}R^{2}\right)+\frac{\eta}{8m^{2}}R^{2}\label{eq:Cubic_theo_equivalent_RaR2}\\
 & +\frac{\alpha}{6m^{4}}\left[R^{\mu\nu}R_{\nu}^{\phantom{\nu}\alpha}R_{\alpha\mu}-\left(1+\frac{\omega}{\sigma\alpha\lambda_{0}}\right)RR_{\mu\nu}^{2}+\gamma R^{3}\right]\Biggr\}\nonumber \end{align}
 that reduces to (\ref{eq:Equiv_RaR2_act}) is unitary or nonunitary.
For concreteness, consider $\eta=0$ and $\omega=+1$, then for $\sigma=-1$,
the theory is not unitary in dS. For $\sigma=+1$, the theory is unitary
if $\xi>\frac{1}{16}$ and $\frac{1}{4\xi}<\lambda_{0}<\frac{1+72\xi+\left(1+48\xi\right)^{3/2}}{864\xi^{2}}$.
In AdS, for $\sigma=+1$, the unitary region is $\xi<0$ and $\lambda_{0}<\frac{1}{4\xi}$.
For $\sigma=-1$, for any value of $\lambda_{0}$ there is a unitary
region for $\xi<0$. The analysis for $\eta\ne0$ can also be done
in the same lines.

\subsection{Central charge and boundary unitarity}

In all the above analysis, we have considered bulk unitarity only.
For the applications of AdS/CFT, boundary unitarity is also relevant.
From the detailed work of \cite{BHT2}, we know that for NMG bulk
and boundary unitarity are in conflict. This conflict is not resolved
in the cubic order extension \cite{Sinha}, or the infinite order
extension of NMG \cite{GulluBINMG,Nam,Paulos,Gullu_cfuncion}. The
bulk and boundary unitarity conflict follows from the requirement
that a positive central charge is not allowed for NMG in the region
where NMG is bulk unitary. Therefore, it would be quite interesting
to find both bulk and boundary unitary higher derivative theories.
As we will see in this section, there are many such theories. First,
recall that the central charge of a generic three-dimensional higher
curvature gravity theory can be found by using \cite{Wald,Kraus,Saida,Imbimbo}
\begin{equation}
c=\frac{8\pi}{\sqrt{\left|\lambda\right|}m}\left[g_{\mu\nu}\frac{\partial\mathcal{L}}{\partial R_{\mu\nu}}\right]_{\bar{R}_{\mu\nu}},\end{equation}
 where the coefficient in front was put to conform to the normalization
of Brown-Henneaux \cite{Brown}. It is easy to see that the central
charge of a generic higher derivative theory can be computed directly
from the equivalent quadratic action, since $\left[\frac{\partial\mathcal{L}}{\partial R_{\mu\nu}}\right]_{\bar{R}_{\mu\nu}}$
is the first order term in the Taylor series expansion of the full
Lagrangian around its constant curvature vacuum. This simple observation
leads to a remarkable conclusion in the light of the discussion above:
\emph{Any higher curvature theory that reduces to NMG cannot be unitary
both in the bulk and on the boundary}. This explains why an extension
of NMG, be it cubic or any power, that has a free theory like NMG
will not have unitarity on the boundary and in the bulk, and hence
perhaps will not be relevant to AdS/CFT. But, \emph{any higher curvature
theory that has the same free theory as the cosmological Einstein
theory} will be unitary both in the bulk and on the boundary. The
theories constructed in Sec.\ref{sub:Reducing-EH} have the central
charge to be \begin{equation}
c=\frac{24\pi\tilde{\sigma}}{\sqrt{\left|\lambda\right|}\kappa^{2}m}.\end{equation}
 Both bulk and boundary unitarity requires $\tilde{\sigma}>0$. But,
these are not the only theories that are unitary everywhere: Let us
now consider the higher curvature theories that have the same free
theory as the $\sigma R-2\Lambda_{0}+aR^{2}$ that we discussed in
Sec.\ref{sub:Reducing-RaR2}. The central charge of (\ref{eq:Equiv_quad_act_of_cubic})
with $\tilde{\omega}=0$ can be computed as \begin{equation}
c=\frac{24\pi}{\sqrt{\left|\lambda\right|}\kappa^{2}m}\left(\tilde{\sigma}+\frac{3\tilde{\eta}\lambda}{2}\right).\end{equation}
 For unitarity $\tilde{\eta}>0$, and in AdS since $\lambda<0$ we
should have $\tilde{\sigma}>-\frac{3\tilde{\eta}\lambda}{2}$ to have
$c>0$. We should check if this constraint is consistent with the
other constraint (the BF bound) $m_{s}^{2}\ge\lambda m^{2}$ with
$m_{s}^{2}=\left(\frac{\tilde{\sigma}}{\tilde{\eta}}-\frac{3\lambda}{2}\right)m^{2}$
and the existence of a negative $\lambda$ satisfying the vacuum equation
$\tilde{\sigma}\lambda-\frac{3\tilde{\eta}}{4}\lambda^{2}=\tilde{\lambda}_{0}$.
One can find families of theories satisfying these bounds, let us
give a simple example for which we take $\eta=0$ and $\omega=1$,
then the action (\ref{eq:Cubic_theo_equivalent_RaR2}) is bulk and
boundary unitary for\begin{align}
 & \sigma=+1\quad\text{and}\quad\xi<0\quad\text{and}\quad\frac{24\xi-\left(1-16\xi\right)^{3/2}-1}{32\xi^{2}}<\lambda_{0}<\frac{1}{4\xi},\nonumber \\
 & \sigma=-1\quad\text{and}\quad-\frac{1}{16}<\xi<0\quad\text{and}\quad-\frac{24\xi+\left(1+16\xi\right)^{3/2}+1}{32\xi^{2}}<\lambda_{0}<-\frac{1}{4\xi},\end{align}
 where $\xi$ was defined just before Sec.\ref{sub:Reducing-NMG}. 

To summarize, if a higher curvature theory is required to be unitary
both in the bulk and on the boundary, then it should have the same
free theory as either the cosmological Einstein-Hilbert theory, or
the $R-2\Lambda_{0}+aR^{2}$ theory with the constraints satisfying
the bounds discussed above.

\section{Unitarity of BINMG}

Up to now, we have constructed all the unitary cubic curvature theories
in (A)dS. The procedure can be carried on to quartic or more powers
of curvature, but here let us give two examples of Born-Infeld gravities
which in principle include infinite powers of curvature. Our first
example is the Born-Infeld extension of NMG was introduced in \cite{GulluBINMG}
with the action \begin{equation}
I_{\text{BINMG}}=-\frac{4m^{2}}{\kappa^{2}}\int d^{3}x\,\left[\sqrt{-\det\left(g+\frac{\sigma}{m^{2}}G\right)}-\left(1-\frac{\lambda_{0}}{2}\right)\sqrt{-\det g}\right],\label{eq:det_BINMG_action}\end{equation}
 where $G_{\mu\nu}\equiv R_{\mu\nu}-\frac{1}{2}g_{\mu\nu}R$ and $\sigma=\pm1$.
This particular form of the action was chosen to reproduce the cosmological
Einstein-Hilbert action at the first order in the curvature expansion
and the NMG in the second order expansion. These two conditions are
actually met by another BI-type action that we shall discuss below
which constitute our second example. On the other hand, the cubic
and fourth order extensions of NMG given in \cite{Sinha} which was
constructed with the help of a holographic $c$-function matches the
same orders of (\ref{eq:det_BINMG_action}). Certain aspects of BINMG
such as its central charge \cite{Nam,Gullu_cfuncion}, $c$-functions
\cite{Gullu_cfuncion}, classical solutions \cite{Nam,Alishahiha,Ghodsi1,Ghodsi2}
have been studied. We will study the unitarity of BINMG with two different
methods: First, with the help of an equivalent quadratic action that
we have employed above, and secondly we will explicitly calculate
the second order expansion in metric perturbation $h_{\mu\nu}$ with
the methods developed in \cite{Gullu_UniBI}. These two methods obviously
will give the same answer, but it is worth checking that the equivalent
quadratic action method works with the help of the second more direct
method for this infinite order theories. This more direct method is
highly involved in terms of computation; therefore, we put it in the
Appendix.

Let us analyze the BINMG action by finding its equivalent quadratic
action: To do that we have to expand the determinant in terms of traces
which was done in \cite{Gullu3} \begin{equation}
I_{\text{BINMG}}=-\frac{4m^{2}}{\kappa^{2}}\int d^{3}x\,\sqrt{-g}\left[\sqrt{1-\frac{\sigma}{2m^{2}}\left(R+\frac{\sigma}{m^{2}}K-\frac{1}{12m^{4}}S\right)}-\left(1-\frac{\lambda_{0}}{2}\right)\right],\label{eq:Expanded_BINMG_action}\end{equation}
 where $K$ and $S$ are defined as\begin{equation}
K\equiv R_{\mu\nu}^{2}-\frac{1}{2}R^{2},\qquad S\equiv8R^{\mu\nu}R_{\mu\alpha}R_{\phantom{\alpha}\nu}^{\alpha}-6RR_{\mu\nu}^{2}+R^{3}.\end{equation}
 The \emph{unique} vacuum of (\ref{eq:Expanded_BINMG_action}) by
directly studying the equations of motion was found in \cite{Gullu_cfuncion,Nam}
as\begin{equation}
\lambda=\sigma\lambda_{0}\left(1-\frac{\lambda_{0}}{4}\right),\quad\lambda_{0}<2.\label{eq:Vacuum_of_BINMG}\end{equation}
 In the spirit of the current work, let us verify this result by finding
the equivalent linear action which circumvents the use of equations
of motion. Let us define\begin{align}
f\left(R_{\nu}^{\mu}\right)\equiv & \left(1-\frac{\sigma}{2m^{2}}\left\{ \delta_{\mu}^{\nu}R_{\nu}^{\mu}+\frac{\sigma}{m^{2}}\left[R_{\nu}^{\mu}R_{\mu}^{\nu}-\frac{1}{2}\left(\delta_{\mu}^{\nu}R_{\nu}^{\mu}\right)^{2}\right]\right.\right.\nonumber \\
 & \left.\left.-\frac{1}{12m^{4}}\left[8R_{\rho}^{\mu}R_{\nu}^{\rho}R_{\mu}^{\nu}-6R_{\nu}^{\mu}R_{\mu}^{\nu}\left(\delta_{\lambda}^{\gamma}R_{\gamma}^{\lambda}\right)+\left(\delta_{\mu}^{\nu}R_{\nu}^{\mu}\right)^{3}\right]\right\} \right)^{1/2}-\left(1-\frac{\lambda_{0}}{2}\right),\end{align}
 which assumes, as above, that $R_{\nu}^{\mu}$ is the independent
variable. Expanding $f\left(R_{\nu}^{\mu}\right)$ around its constant
curvature background $\left(\bar{R}_{\nu}^{\mu}=2\lambda m^{2}\delta_{\nu}^{\mu}\right)$
to the first order in $\left(R_{\alpha}^{\beta}-\bar{R}_{\alpha}^{\beta}\right)$
as (\ref{eq:First_order_in_R-Rbar}) one can find the equivalent linear
Lagrangian density. For this one needs \begin{equation}
f\left(\bar{R}_{\nu}^{\mu}\right)=\left(1-\sigma\lambda\right)^{3/2}-\left(1-\frac{\lambda_{0}}{2}\right),\label{eq:Value_of_action}\end{equation}
 which requires $\sigma\lambda\le1$, \begin{align}
\left[\frac{\partial f}{\partial R_{\beta}^{\alpha}}\right]_{\bar{R}_{\nu}^{\mu}}= & -\frac{\sigma}{4m^{2}\left(1-\sigma\lambda\right)^{3/2}}\nonumber \\
 & \left[\delta_{\alpha}^{\beta}+\frac{\sigma}{m^{2}}\left(2\bar{R}_{\alpha}^{\beta}-\bar{R}\delta_{\alpha}^{\beta}\right)-\frac{1}{12m^{4}}\left(24\bar{R}_{\lambda}^{\beta}\bar{R}_{\alpha}^{\lambda}-12\bar{R}_{\alpha}^{\beta}\bar{R}-6\bar{R}_{\lambda}^{\gamma}\bar{R}_{\gamma}^{\lambda}\delta_{\alpha}^{\beta}+3\bar{R}^{2}\delta_{\alpha}^{\beta}\right)\right]\end{align}
 which requires $\sigma\lambda\ne1$, then\begin{align}
\left[\frac{\partial f}{\partial R_{\beta}^{\alpha}}\right]_{\bar{R}_{\nu}^{\mu}} & =-\frac{\sigma\delta_{\alpha}^{\beta}}{4m^{2}}\sqrt{1-\sigma\lambda}.\label{eq:First_derivative}\end{align}
 With these results, the equivalent linear action for BINMG becomes\begin{equation}
I_{\text{lin-equal}}=\frac{\sigma\sqrt{1-\sigma\lambda}}{\kappa^{2}}\int d^{3}x\,\sqrt{-g}\left\{ R-4\sigma m^{2}\left[1+\frac{\sigma}{2}\lambda+\frac{1}{2\sqrt{1-\sigma\lambda}}\left(\lambda_{0}-2\right)\right]\right\} ,\end{equation}
 where one can read the effective cosmological constant as\begin{align}
\lambda & =2\sigma\left[1+\frac{\sigma}{2}\lambda+\frac{1}{2\sqrt{1-\sigma\lambda}}\left(\lambda_{0}-2\right)\right]\Rightarrow2\sqrt{1-\sigma\lambda}=2-\lambda_{0},\end{align}
 which requires $\lambda_{0}<2$, and after taking the square of the
equation, one obtains (\ref{eq:Vacuum_of_BINMG}).

Expansion of $f\left(R_{\nu}^{\mu}\right)$ around the constant curvature
background by using (\ref{eq:Hindawi_exp}) with the assumption of
small fluctuations about the background requires the quantity\begin{align}
\left[\frac{\partial^{2}f}{\partial R_{\sigma}^{\rho}\partial R_{\beta}^{\alpha}}\right]_{\bar{R}_{\nu}^{\mu}}= & -\frac{1}{2m^{4}\sqrt{1-\sigma\lambda}}\left(\delta_{\rho}^{\beta}\delta_{\alpha}^{\sigma}-\frac{3}{8}\delta_{\alpha}^{\beta}\delta_{\rho}^{\sigma}\right).\end{align}
 Using this and (\ref{eq:Value_of_action}), (\ref{eq:First_derivative});
one obtains the equivalent quadratic action as\begin{equation}
I_{O\left(R^{2}\right)}=\frac{1}{\kappa^{2}}\int d^{3}x\sqrt{-g}\,\left[\tilde{\sigma}R-2m^{2}\tilde{\lambda}_{0}+\frac{\tilde{\omega}}{m^{2}}\left(R_{\mu\nu}^{2}-\frac{3}{8}R^{2}\right)\right],\label{eq:Equivalent_quadratic_of_BINMG}\end{equation}
 where, for $\sigma\lambda<1$,\begin{equation}
\tilde{\sigma}=\frac{\left(\sigma-\frac{\lambda}{2}\right)}{\sqrt{1-\sigma\lambda}},\qquad\tilde{\lambda}_{0}=\lambda_{0}-2+\frac{1}{\sqrt{1-\sigma\lambda}}\left(2-\sigma\lambda-\frac{\lambda^{2}}{4}\right),\qquad\tilde{\omega}=\frac{1}{\sqrt{1-\sigma\lambda}}.\label{eq:Parameters_of_equiv_quad}\end{equation}
 Remarkably, the equivalent quadratic action turned out to be NMG
with redefined parameters. Namely, the effect of all the terms beyond
$O\left(R^{2}\right)$ simply change the parameters of the $O\left(R^{2}\right)$
expansion of the action which was NMG by construction. Let us stress
again that this equivalent quadratic action has the same free theory
(that is the propagator), same vacuum and same central charge as BINMG.
Vacuum of BINMG in terms of the redefined parameters is \begin{equation}
\tilde{\omega}\lambda^{2}+4\tilde{\sigma}\lambda-4\tilde{\lambda}_{0}=0.\label{eq:Vacuum_of_BINMG_in_tilde_params}\end{equation}
 From the discussion in Sec.\ref{sub:Reducing-NMG}, we know that
NMG is unitary under two conditions $\tilde{\omega}\lambda-2\tilde{\sigma}>0$
and $\frac{2\tilde{\sigma}}{\tilde{\omega}}+\lambda\le0$. Now, the
question is whether these conditions are satisfied together with the
BINMG condition $\lambda_{0}<2$ or not. A simple analysis shows that
BINMG is unitary only for $\sigma=-1$ in AdS for $0<\lambda_{0}<2$,
and in dS for $\lambda_{0}<0$. Therefore, this analysis answers the
question raised in \cite{Gullu_cfuncion} about the unitarity of the
$\sigma=+1$ theory in the negative. This is true for bulk unitarity,
for boundary unitarity recall the central charge from \cite{Nam,Gullu_cfuncion},
or just compute it from the equivalent action (\ref{eq:Equivalent_quadratic_of_BINMG})
as\begin{equation}
c=\frac{3\ell}{2G_{3}}\left(\tilde{\sigma}-\frac{\tilde{\omega}\lambda}{2}\right)=\frac{3\sigma\ell}{4G_{3}}\left(2-\lambda_{0}\right).\end{equation}
 Since in AdS $0<\lambda_{0}<2$, and $\sigma=-1$, the theory is
not unitary on the boundary just like NMG, or the cubic extension
of NMG. The $\sigma=+1$ theory is unitary on the boundary, but as
we have just seen it is not unitary in the bulk. This is an expected
result, because the free theory of BINMG is the same as the free theory
of NMG with redefined parameters, and there is the obvious conflict
between the bulk unitarity condition $\tilde{\omega}\lambda-2\tilde{\sigma}>0$
and the boundary unitarity condition $2\tilde{\sigma}-\tilde{\omega}\lambda>0$.

We mentioned that there was a second BI-type action that reproduces
NMG in the curvature expansion. The action of this theory reads \cite{GulluBINMG}
\begin{equation}
I=-\frac{4m^{2}}{\kappa^{2}}\int d^{3}x\,\left\{ \sqrt{-\det\left[g_{\mu\nu}+\frac{\sigma}{m^{2}}\left(R_{\mu\nu}-\frac{1}{6}g_{\mu\nu}R\right)\right]}-\left(1-\frac{\lambda_{0}}{2}\right)\sqrt{-\det g}\right\} ,\end{equation}
 which, by use of \begin{equation}
\det A=\frac{1}{6}\left[\left(\text{Tr}A\right)^{3}-3\text{Tr}A\text{Tr}\left(A^{2}\right)+2\text{Tr}\left(A^{3}\right)\right],\end{equation}
 becomes\begin{align}
I & =-\frac{4m^{2}}{\kappa^{2}}\int d^{3}x\,\sqrt{-g}\nonumber \\
 & \times\left\{ \sqrt{1+\frac{1}{2m^{2}}\left[R-\frac{1}{m^{2}}\left(R_{\mu\nu}^{2}-\frac{1}{2}R^{2}\right)+\frac{2}{3m^{4}}\left(R^{\mu\nu}R_{\nu}^{\phantom{\nu}\alpha}R_{\alpha\mu}-\frac{5}{4}RR_{\mu\nu}^{2}+\frac{23}{72}R^{3}\right)\right]}-\left(1-\frac{\lambda_{0}}{2}\right)\right\} .\end{align}
 Quite interestingly, this action reduces to NMG at $O\left(h^{2}\right)$
with the same redefined parameters as the BINMG. Therefore, at the
free level, these two theories cannot be distinguished.

\section{Conclusion}

We have found all the unitary cubic curvature theories in three dimensions
around constant curvature backgrounds. Without any further constraint,
we have shown that unitarity in the bulk and on the boundary allows
a large family of solutions as opposed to the cubic curvature theories
that have appeared in the literature before, which allowed only bulk
or boundary unitarity. The theories we have found should be studied
in the context of AdS/CFT. We have also studied the unitarity of two
Born-Infeld extensions of NMG which turned out to be unitary in the
bulk only. Besides the parity violating extension with the addition
of a Chern-Simons term and/or carrying out the unitarity analysis
to $O\left(R^{4}\right)$, a quite physically relevant extension of
our work is to find the unitary cubic curvature theories in four dimensions,
which is currently under construction.

\section{Acknowledgments}

This work is supported by the T{Ü}B\.{I}TAK Grant No. 110T339, and
METU Grant BAP-07-02-2010-00-02. Some of the calculations in this
paper were either done or checked with the help of the computer package
Cadabra \cite{Cadabra1,Cadabra2}.

\section*{Appendix: $O\left(h^{2}\right)$ Action of BINMG }

In this Appendix, we calculate explicitly $O\left(h\right)$ and $O\left(h^{2}\right)$
expansions of the BINMG action. First of all, let us find the constant
curvature vacuum of (\ref{eq:det_BINMG_action}) by explicitly calculating
the first order action in the metric perturbation. In \cite{Gullu_UniBI},
it was shown that $O\left(h\right)$ of the generic BI-type action\begin{equation}
I=\frac{2}{\kappa\alpha}\int d^{D}x\,\left[\sqrt{-\det\left(g_{\mu\nu}+A_{\mu\nu}\right)}-\left(\alpha\Lambda_{0}+1\right)\sqrt{-\det g}\right],\label{eq:Generic_BI_action}\end{equation}
 where $A_{\mu\nu}$ is in the form $A_{\mu\nu}=\alpha\left(R_{\mu\nu}+\beta\tilde{R}_{\mu\nu}\right)+O\left(R^{2}\right)$
with the definition $\tilde{R}_{\mu\nu}\equiv R_{\mu\nu}-\frac{1}{D}g_{\mu\nu}R$
is\begin{equation}
I_{O\left(h\right)}=\frac{\left(1+a\right)^{\frac{D-4}{2}}}{\kappa\alpha}\int d^{D}x\,\sqrt{-\bar{g}}\left[\left(1+a\right)\left(\bar{g}^{\rho\mu}A_{\mu\rho}^{\left(1\right)}+h\right)-\left(1+a\right)^{\frac{4-D}{2}}\left(\alpha\Lambda_{0}+1\right)h\right],\label{eq:Oh_action}\end{equation}
 where $A_{\mu\rho}^{\left(1\right)}$ is the first order term in
the metric perturbation expansion of $A_{\mu\nu}$. Here, $a$ is
defined as $\bar{A}_{\mu\nu}\equiv a\bar{g}_{\mu\nu}$ and for BINMG
it becomes\begin{equation}
\frac{\sigma}{m^{2}}\left(\bar{R}_{\mu\nu}-\frac{1}{2}\bar{g}_{\mu\nu}\bar{R}\right)=-\sigma\lambda\bar{g}_{\mu\nu}\Rightarrow a=-\sigma\lambda,\end{equation}
 which, when inserted to the action, yields the constraint $a>-1\Rightarrow\sigma\lambda<1$.
For BINMG, $A_{\mu\nu}$ is $A_{\mu\nu}=\frac{\sigma}{m^{2}}\left(R_{\mu\nu}-\frac{1}{2}g_{\mu\nu}R\right)$,
then $A_{\mu\nu}^{\left(1\right)}$ and $\bar{g}^{\rho\mu}A_{\mu\rho}^{\left(1\right)}$
becomes\begin{equation}
A_{\mu\nu}^{\left(1\right)}=\frac{\sigma}{m^{2}}\left(R_{\mu\nu}^{L}-\frac{1}{2}\bar{g}_{\mu\nu}R_{L}-3\lambda m^{2}h_{\mu\nu}\right),\qquad\bar{g}^{\rho\mu}A_{\mu\rho}^{\left(1\right)}=-\frac{\sigma}{2m^{2}}\left(R_{L}+2\lambda m^{2}h\right),\label{eq:Linearized_A-tensor_terms}\end{equation}
 where $R_{\mu\nu}^{L}$ and $R_{L}$ are the linearized Ricci tensor
and the linearized curvature scalar with the definitions\begin{equation}
R_{\mu\nu}^{L}\equiv\frac{1}{2}\left(\bar{\nabla}_{\sigma}\bar{\nabla}_{\mu}h_{\nu}^{\sigma}+\bar{\nabla}_{\sigma}\bar{\nabla}_{\nu}h_{\mu}^{\sigma}-\bar{\Box}h_{\mu\nu}-\bar{\nabla}_{\mu}\bar{\nabla}_{\nu}h\right),\qquad R_{L}\equiv\left(g^{\mu\nu}R_{\mu\nu}\right)_{L}.\end{equation}
 Then, for BINMG with $\alpha=-\frac{1}{2m^{2}}$ and $\kappa\rightarrow\kappa^{2}$,
the $O\left(h\right)$ action becomes\begin{align}
I_{O\left(h\right)} & =-\frac{2m^{2}}{\kappa^{2}\sqrt{1+a}}\int d^{3}x\,\sqrt{-\bar{g}}\left\{ \left(1+a\right)\left[-\frac{\sigma}{2m^{2}}\left(R_{L}+2\lambda m^{2}h\right)\right]+\left(1+a\right)h-\sqrt{1+a}\left(1-\frac{\lambda_{0}}{2}\right)h\right\} \nonumber \\
 & =-\frac{2m^{2}}{\kappa^{2}\sqrt{1+a}}\int d^{3}x\,\sqrt{-\bar{g}}\left[\left(1+a\right)h-\sqrt{1+a}\left(1-\frac{\lambda}{2}\right)h\right],\end{align}
 then the constant curvature background equation of motion can be
found as in (\ref{eq:Vacuum_of_BINMG}) from the coefficient of $h^{\mu\nu}$.

Now, let us turn to the explicit calculation of $O\left(h^{2}\right)$
action for BINMG. In \cite{Gullu_UniBI}, the second order action
in metric perturbation for (\ref{eq:Generic_BI_action}) in three
dimensions was calculated as\begin{align}
I_{O\left(h^{2}\right)}= & -\frac{1}{\kappa\alpha\sqrt{1+a}}\int d^{3}x\,\sqrt{-\bar{g}}\left\{ \frac{1}{2}A_{\mu\nu}^{\left(1\right)}A_{\left(1\right)}^{\mu\nu}-\frac{1}{4}\left(\bar{g}^{\mu\nu}A_{\mu\nu}^{\left(1\right)}\right)^{2}-\left(1+a\right)\bar{g}^{\mu\nu}A_{\mu\nu}^{\left(2\right)}\right.\nonumber \\
 & \left.+h^{\mu\nu}\left(A_{\mu\nu}^{\left(1\right)}-\frac{1}{2}\bar{g}_{\mu\nu}\bar{g}^{\rho\sigma}A_{\rho\sigma}^{\left(1\right)}\right)-\frac{1}{4}\left[1-\sqrt{1+a}\left(\alpha\Lambda_{0}+1\right)\right]\left(h^{2}-2h_{\mu\nu}^{2}\right)\right\} .\label{eq:Oh2_action_in_3D}\end{align}
 With the explicit form of $A_{\mu\nu}$ for BINMG, let us calculate
each term separately. First, the second line of the above equation
takes the following form by use of the definition of the linearized
Einstein tensor $\mathcal{G}_{\mu\nu}^{L}\equiv R_{\mu\nu}^{L}-\frac{1}{2}\bar{g}_{\mu\nu}R_{L}-2\Lambda h_{\mu\nu}$
in three dimensions and by use of the equation of motion;\begin{multline}
\int d^{3}x\,\sqrt{-\bar{g}}\left\{ h^{\mu\nu}\left(A_{\mu\nu}^{\left(1\right)}-\frac{1}{2}\bar{g}_{\mu\nu}A_{\alpha}^{\alpha\left(1\right)}\right)-\frac{1}{4}\left[1-\sqrt{1+a}\left(1-\frac{\lambda_{0}}{2}\right)\right]h^{\mu\nu}\left(\bar{g}_{\mu\nu}h-2h_{\mu\nu}\right)\right\} \\
=\frac{\sigma}{m^{2}}\int d^{3}x\,\sqrt{-\bar{g}}h^{\mu\nu}\left[\mathcal{G}_{\mu\nu}^{L}+\frac{1}{4}\bar{g}_{\mu\nu}R_{L}+\frac{\lambda m^{2}}{4}\left(\bar{g}_{\mu\nu}h-2h_{\mu\nu}\right)\right].\end{multline}
 Secondly, let us calculate the terms quadratic in $A_{\mu\nu}$.
There are two such terms $A_{\mu\nu}^{\left(1\right)}A_{\left(1\right)}^{\mu\nu}$
and $\left(\bar{g}^{\mu\nu}A_{\mu\nu}^{\left(1\right)}\right)^{2}$,
and the first one becomes \begin{align}
\int d^{3}x\,\sqrt{-\bar{g}}\frac{1}{2}A_{\mu\nu}^{\left(1\right)}A_{\left(1\right)}^{\mu\nu}=\frac{1}{2m^{4}}\int d^{3}x\,\sqrt{-\bar{g}}h^{\mu\nu} & \biggl[-\frac{1}{4}\left(\bar{g}_{\mu\nu}\bar{\Box}-\bar{\nabla}_{\mu}\bar{\nabla}_{\nu}+2\lambda m^{2}\bar{g}_{\mu\nu}\right)R_{L}\nonumber \\
 & -\frac{1}{2}\left(\bar{\Box}\mathcal{G}_{\mu\nu}^{L}-\lambda m^{2}\bar{g}_{\mu\nu}R_{L}\right)-\lambda m^{2}\mathcal{G}_{\mu\nu}^{L}+\lambda^{2}m^{4}h_{\mu\nu}\biggr],\end{align}
 by using $\int d^{3}x\,\sqrt{-\bar{g}}R_{L}^{2}$ and $\int d^{3}x\,\sqrt{-\bar{g}}R_{L}^{\mu\nu}R_{\mu\nu}^{L}$
which can be found as \begin{equation}
\int d^{3}x\,\sqrt{-\bar{g}}R_{L}^{2}=\int d^{3}x\,\sqrt{-\bar{g}}\left[-h^{\mu\nu}\left(\bar{g}_{\mu\nu}\bar{\Box}-\bar{\nabla}_{\mu}\bar{\nabla}_{\nu}+2\lambda m^{2}\bar{g}_{\mu\nu}\right)R_{L}\right],\end{equation}
 \begin{align}
\int d^{3}x\,\sqrt{-\bar{g}}R_{L}^{\mu\nu}R_{\mu\nu}^{L}=-\frac{1}{2}\int d^{3}x\,\sqrt{-\bar{g}}h^{\mu\nu} & \Biggl[\left(\bar{g}_{\mu\nu}\bar{\Box}-\bar{\nabla}_{\mu}\bar{\nabla}_{\nu}+2\lambda m^{2}\bar{g}_{\mu\nu}\right)R_{L}+\left(\bar{\Box}\mathcal{G}_{\mu\nu}^{L}-\lambda m^{2}\bar{g}_{\mu\nu}R_{L}\right)\nonumber \\
 & -10\lambda m^{2}R_{\mu\nu}^{L}+\lambda m^{2}\bar{g}_{\mu\nu}R_{L}+12\lambda^{2}m^{4}h_{\mu\nu}\Biggr],\end{align}
 where the background Bianchi identity and integration by parts have
been used. The other term reads \begin{align}
\int d^{3}x\,\sqrt{-\bar{g}}\left[-\frac{1}{4}\left(A_{\alpha}^{\alpha\left(1\right)}\right)^{2}\right]=\frac{1}{16m^{4}}\int d^{3}x\,\sqrt{-\bar{g}}h^{\mu\nu} & \biggl[\left(\bar{g}_{\mu\nu}\bar{\Box}-\bar{\nabla}_{\mu}\bar{\nabla}_{\nu}+2\lambda m^{2}\bar{g}_{\mu\nu}\right)R_{L}\nonumber \\
 & -4\lambda m^{2}\bar{g}_{\mu\nu}R_{L}-4\lambda^{2}m^{4}\bar{g}_{\mu\nu}h\biggr].\end{align}
 Let us consider $\bar{g}^{\mu\nu}A_{\mu\nu}^{\left(2\right)}$ which
is ,\begin{align}
\bar{g}^{\mu\nu}A_{\mu\nu}^{\left(2\right)}= & \frac{\sigma}{m^{2}}\left(\bar{g}^{\mu\nu}R_{\mu\nu}^{\left(2\right)}-\frac{3}{2}R^{\left(2\right)}-\frac{1}{2}hR_{L}\right),\end{align}
 and using \begin{equation}
\int d^{3}x\,\sqrt{-\bar{g}}R_{\left(2\right)}=\int d^{3}x\,\sqrt{-\bar{g}}h^{\mu\nu}\left(-\frac{1}{2}\mathcal{G}_{\mu\nu}^{L}-\frac{1}{2}\bar{g}_{\mu\nu}R_{L}+\lambda m^{2}h_{\mu\nu}-\frac{\lambda m^{2}}{2}\bar{g}_{\mu\nu}h\right),\label{eq:Integral_R2}\end{equation}
 and \begin{equation}
\int d^{3}x\,\sqrt{-\bar{g}}\bar{g}^{\mu\nu}R_{\mu\nu}^{\left(2\right)}=h^{\mu\nu}\left(\frac{1}{2}\mathcal{G}_{\mu\nu}^{L}+\lambda m^{2}h_{\mu\nu}-\frac{\lambda m^{2}}{2}\bar{g}_{\mu\nu}h\right),\end{equation}
 one gets\begin{equation}
\int d^{3}x\,\sqrt{-\bar{g}}\bar{g}^{\mu\nu}A_{\mu\nu}^{\left(2\right)}=\int d^{3}x\,\sqrt{-\bar{g}}h^{\mu\nu}\left[\frac{\sigma}{4m^{2}}\left(5\mathcal{G}_{\mu\nu}^{L}+\bar{g}_{\mu\nu}R_{L}+\lambda m^{2}\bar{g}_{\mu\nu}h-2\lambda m^{2}h_{\mu\nu}\right)\right].\end{equation}
 This computation is somewhat lengthy, and one needs\begin{equation}
\bar{g}^{\nu\sigma}h_{\beta}^{\mu}\left(R_{\phantom{\mu}\nu\mu\sigma}^{\beta}\right)_{L}=h^{\mu\nu}\left(-R_{\mu\nu}^{L}+3\lambda m^{2}h_{\mu\nu}-\lambda m^{2}\bar{g}_{\mu\nu}h\right),\label{eq:ghRiemann}\end{equation}
 and the two expressions involving linearized Christoffel connection
whose definition is $\left(\Gamma_{\mu\nu}^{\rho}\right)_{L}\equiv\frac{1}{2}\bar{g}^{\rho\lambda}\left(\bar{\nabla}_{\mu}h_{\nu\lambda}+\bar{\nabla}_{\nu}h_{\mu\lambda}-\bar{\nabla}_{\lambda}h_{\mu\nu}\right)$,
\begin{align}
\int d^{3}x\,\sqrt{-\bar{g}}\bar{g}^{\nu\sigma}\bar{g}^{\mu\alpha}\bar{g}_{\beta\gamma}\left(\Gamma_{\mu\alpha}^{\gamma}\right)_{L}\left(\Gamma_{\sigma\nu}^{\beta}\right)_{L} & =\int d^{3}x\,\sqrt{-\bar{g}}\left[-\frac{1}{2}h^{\mu\nu}\left(\bar{\nabla}^{\sigma}\bar{\nabla}_{\mu}h_{\nu\sigma}+\bar{\nabla}^{\sigma}\bar{\nabla}_{\nu}h_{\mu\sigma}-\frac{3}{2}\bar{\nabla}_{\mu}\bar{\nabla}_{\nu}h\right)\right.\nonumber \\
 & \phantom{=\int d^{3}x\,\sqrt{-\bar{g}}}\left.+h^{\mu\nu}\left(3\lambda m^{2}h_{\mu\nu}-\frac{\lambda m^{2}}{2}\bar{g}_{\mu\nu}h\right)+\frac{1}{4}h^{\mu\nu}\bar{g}_{\mu\nu}R_{L}\right],\end{align}
 \begin{equation}
\int d^{3}x\,\sqrt{-\bar{g}}\bar{g}^{\nu\sigma}\bar{g}^{\mu\alpha}\bar{g}_{\beta\gamma}\left(\Gamma_{\sigma\alpha}^{\gamma}\right)_{L}\left(\Gamma_{\mu\nu}^{\beta}\right)_{L}=\int d^{3}x\,\sqrt{-\bar{g}}\left[-\frac{3}{4}h^{\mu\nu}\bar{\Box}h_{\mu\nu}+\frac{1}{4}h^{\mu\nu}\left(\bar{\nabla}^{\sigma}\bar{\nabla}_{\mu}h_{\nu\sigma}+\bar{\nabla}^{\sigma}\bar{\nabla}_{\nu}h_{\mu\sigma}\right)\right].\end{equation}
 Collecting all the terms and making use of the equations of motion,
one obtains

\begin{align}
I_{O\left(h^{2}\right)} & =-\frac{1}{2\kappa^{2}\sqrt{1-\sigma\lambda}}\int d^{3}x\,\sqrt{-\bar{g}}h^{\mu\nu}\label{eq:BINMG_linear}\\
 & \times\left\{ \left(\sigma-3\lambda\right)\mathcal{G}_{\mu\nu}^{L}+\frac{1}{m^{2}}\left[\frac{1}{4}\left(\bar{g}_{\mu\nu}\bar{\Box}-\bar{\nabla}_{\mu}\bar{\nabla}_{\nu}+2\lambda m^{2}\bar{g}_{\mu\nu}\right)R_{L}+\left(\bar{\Box}\mathcal{G}_{\mu\nu}^{L}-\lambda m^{2}\bar{g}_{\mu\nu}R_{L}\right)\right]\right\} ,\nonumber \end{align}
 which can be compared to (25) of \cite{DeserTekin}. Then, one can
observe that this is the $O\left(h^{2}\right)$ of NMG with the redefined
parameters given in (\ref{eq:Parameters_of_equiv_quad}).


\begin{thebibliography}{37}
\bibitem{BHT1} E.~A.~Bergshoeff, O.~Hohm and P.~K.~Townsend,
Phys.\ Rev.\ Lett.\ \textbf{102}, 201301 (2009).

\bibitem{BHT2} E.~A.~Bergshoeff, O.~Hohm and P.~K.~Townsend,
Phys.\ Rev.\ D\textbf{ 79}, 124042 (2009). 

\bibitem{Gullu_UniBI} I.~Gullu, T.~C.~Sisman and B.~Tekin, Phys.\ Rev.\ D
\textbf{82}, 124023 (2010). 

\bibitem{Hindawi} A.~Hindawi, B.~A.~Ovrut and D.~Waldram, Phys.\ Rev.\ D
\textbf{53}, 5597 (1996).

\bibitem{Tseytlin} R.~R.~Metsaev and A.~A.~Tseytlin, Phys.\ Lett.\ B
\textbf{185}, 52 (1987).

\bibitem{Niedermaier} M.~Niedermaier, Class.\ Quant.\ Grav.\ \textbf{24},
R171 (2007). 

\bibitem{Sinha} A.~Sinha, JHEP \textbf{1006}, 061 (2010); arXiv:1008.4315
{[}hep-th{]}. 

\bibitem{GulluBINMG} I.~Gullu, T.~C.~Sisman and B.~Tekin, Class.\ Quant.\ Grav.\ \textbf{27},
162001 (2010).

\bibitem{Gullu_cfuncion} I.~Gullu, T.~C.~Sisman and B.~Tekin,
Phys.\ Rev.\ D \textbf{82}, 024032 (2010).

\bibitem{Paulos} M.~F.~Paulos, Phys.\ Rev.\ D \textbf{82}, 084042
(2010).

\bibitem{Nakasone} M.~Nakasone and I.~Oda, Prog.\ Theor.\ Phys.
\textbf{121}, 1389 (2009).

\bibitem{Deser} S.~Deser, Phys.\ Rev.\ Lett.\ \textbf{103}, 101302
(2009).

\bibitem{Gullu1} I.~Gullu and B.~Tekin, Phys.\ Rev.\ D \textbf{80},
064033 (2009).

\bibitem{Gullu2} I.~Gullu, T.~C.~Sisman and B.~Tekin, Phys.\ Rev.\ D\textbf{
81}, 104017 (2010).

\bibitem{Blagojevic} M.~Blagojevic and B.~Cvetkovic, arXiv:1010.2596
{[}gr-qc{]}.

\bibitem{Higuchi} A.~Higuchi, Nucl.\ Phys.\ B \textbf{282}, 397
(1987). 

\bibitem{BF} P.~Breitenlohner and D.~Z.~Freedman, Phys.\ Lett.\ B
\textbf{115}, 197 (1982). 

\bibitem{Waldron} A.~R.~Gover, A.~Shaukat and A.~Waldron, Nucl.\ Phys.\ B
\textbf{812}, 424 (2009).

\bibitem{Nepomechie} S.~Deser and R.~I.~Nepomechie, Annals Phys.\ \textbf{154},
396 (1984).

\bibitem{DeserWaldron} S.~Deser and A.~Waldron, Nucl.\ Phys.\ B
\textbf{607}, 577 (2001).

\bibitem{Tekin} B.~Tekin, arXiv:hep-th/0306178. 

\bibitem{Andringa} R.~Andringa, E.~A.~Bergshoeff, M.~de Roo,
O.~Hohm, E.~Sezgin and P.~K.~Townsend, Class.\ Quant.\ Grav.\ \textbf{27},
025010 (2010). 

\bibitem{DeserTekin} S.~Deser and B.~Tekin, Phys.\ Rev.\ D \textbf{67},
084009 (2003).

\bibitem{Stelle} K.~S.~Stelle, Phys.\ Rev.\ D\textbf{ 16}, 953
(1977).

\bibitem{Gullu} I.~Gullu, T.~C.~Sisman and B.~Tekin, Phys.\ Rev.\ D
\textbf{82}, 024032 (2010). 

\bibitem{Nam} S.~Nam, J.~D.~Park and S.~H.~Yi, JHEP \textbf{1007},
058 (2010).

\bibitem{Alishahiha} M.~Alishahiha, A.~Naseh and H.~Soltanpanahi,
Phys.\ Rev.\ D \textbf{82}, 024042 (2010).

\bibitem{Ghodsi1} A.~Ghodsi and M.~Moghadassi, arXiv:1007.4323
{[}hep-th{]}. 

\bibitem{Ghodsi2} A.~Ghodsi and D.~M.~Yekta, arXiv:1010.2434 {[}hep-th{]}. 

\bibitem{Gullu3} I.~Gullu, T.~C.~Sisman and B.~Tekin, Phys.\ Rev.\ D\textbf{
81}, 104018 (2010).

\bibitem{Wald} R.~M.~Wald, Phys.\ Rev.\ D \textbf{48}, R3427
(1993).

\bibitem{Kraus} P.~Kraus and F.~Larsen, JHEP \textbf{0509}, 034
(2005).

\bibitem{Saida} H.~Saida and J.~Soda, Phys.\ Lett.\ B \textbf{471},
358 (2000).

\bibitem{Imbimbo} C.~Imbimbo, A.~Schwimmer, S.~Theisen and S.~Yankielowicz,
Class.\ Quant.\ Grav.\ \textbf{17}, 1129 (2000).

\bibitem{Brown} J.~D.~Brown and M.~Henneaux, Commun.\ Math.\ Phys.\ \textbf{104},
207 (1986).

\bibitem{Cadabra1} K.~Peeters, Comput.\ Phys.\ Commun.\ \textbf{176},
550 (2007).

\bibitem{Cadabra2} K.~Peeters, {}``Introducing Cadabra: A symbolic
computer algebra system for field theory problems'', arXiv:hep-th/0701238.
\end{thebibliography}
\end{document}